\RequirePackage{fix-cm}
\documentclass{svjour3}                     
\smartqed  
\usepackage{graphicx}
\usepackage{url}
\usepackage{lineno}
\usepackage{soul}
\usepackage{amsmath}
\usepackage{amssymb}
\newcommand{\m}[1]{#1\,\text{m}}

\newcommand{\mm}[1]{#1\,\text{mm}}

\begin{document}

\title{A method to estimate particle sizes using the open source software OpenPTV.
}

\titlerunning{Method to estimate sizes with OpenPTV.}        

\author{R.G. Ramirez de la Torre         
        \and
        Atle Jensen
}


\institute{R.G. Ramirez de la Torre \at
             University if Oslo \\
              \email{reynar@math.uio.no}           
           \and
           A. Jensen \at
              University of Oslo
}

\date{Received: date / Accepted: date}

\maketitle

\begin{abstract}
A method to obtain particle sizes from images that are  used for particle tracking velocimetry is proposed. This is an open source method, developed to use together with the open source software OpenPTV. First, the analysis of different factors that affect the estimation of the particle size is made. Then, a transformation is proposed to estimate the sizes from information obtained in OpenPTV. The method requires an extra calibration with a flat target containing circles of different sizes, but if the calibration is successful, the sizes of particles can be estimated reliable within certain limits. For example, for solid of non-translucent particles, sizes can be estimated with an error of maximum 10 \%, as long as the particle diameter correspond to at least three pixels width in the obtained images. For translucent particles some extra assumptions will be needed in the transformation.
\end{abstract}

\section{Introduction}
\label{intro}
In different sciences, the distribution of sizes of observed objects or particles can be important, for example, in the study of multi phase flows. The distribution of the sizes can also be important when estimating source functions of droplets created in the ocean's boundary layer. 
In some cases it is also important to combine measurements of the size of particles together with their kinematics, in that way information of the flow that surrounds them can also be obtained. In this case, techniques that follow the particles in a flow, can provide the Lagrangian description of the surrounding flow. Using holography technique, detailed study of size and relative position of the particles can be done, from where tracking of the trajectories can also be done \cite{katz2010applications}. Laser based methods also allow the measurement of velocity of the particle simultaneously to the size measurements \cite{black1996laser,bachalo1984phase,koothur2021tracking,damaschke2002optical}. The typical accuracy of the optical techniques range between 5 and 15\% \cite{black1996laser,tropea2011optical,kozul2019scanning}.
Other alternatives to laser or holography can also be explored. Specially in the case where the experimental setup or location does not allow the use of these techniques and only simple imaging is available.

In later years open source software has become very popular in the scientific community. Open source software has the advantage that the source code and interfaces are made available for the community to explore, modify and distribute. This is important because it facilitates the sharing of the ideas, reduces the costs and time necessary to create a new software with the same characteristics and functionality. In addition, new features can be added and build upon a common platform, which facilitates the reuse of previous codes rather than individual in-house development, which is probably redundant throughout different research groups. 
In this work, the open source software OpenPTV was used. 
OpenPTV is the abbreviation for the Open Source Particle Tracking Velocimetry consortium \cite{openptv2014openptv}. The OpenPTV foundation is a collaborative effort of several research groups to develop a software for Particle Tracking Velocimetry in 3 Dimensions (3DPTV). The method consists in basically three steps: the identification of illuminated tracer particles from multiple camera views, a triangulation of the probable 3D location of the particles, and the link of particle locations in time to form probable particle trajectories. This technique is well-established and have been used for the study of 3D Lagrangian particle motion in turbulent flows \cite{maas1993particle,malik1993particle,luthi2005lagrangian}.
To reconstruct particle trajectories, the accurate triangulation of the locations of tracer particles, and the unambiguous link of particles to form trajectories are needed. 
The goal of establishing such particle tracks is often to calculate Lagrangian velocities and accelerations. Today there exists different software with similar algorithms to perform PTV. But none of them present an extra feature where the sizes of the particles can be analyzed simultaneously. 

In the present work, a methodology to find an estimate of the size of particles in addition to the tracking of their trajectories is presented. The calibration procedure and the technique are detailed in the following. The code uses Python as primary coding language and is available to complement the use of the open source software OpenPTV. The sizing not only complements the tracking by PTV but depends also on the quality of the tracking. Some of the error sources are analyzed an presented. This is a first approximation of the procedure which has proved to give large improvements by reducing the error.


\section{Methodology}
\label{sec:1}
Particle tracking has shown to be a reliable technique to estimate kinematics of particles moving in a flow. When combined with stereoscopic camera vision, it can achieve 3 dimensional tracing of the particles. The uncertainties in the positions of the particles depend primarily on the optical properties of the media involved, the imaging arrangement and quality of the images, and the quality of the calibration \cite{dracos1996particle}. Therefore, to obtain a successful tracking setup, it is necessary to perform a rigorous and high quality calibration. This calibration, when successful, should allow to obtain an accurate pixel-to-world transformation that can be applied on the retrieved images. Moreover, by using the results of the PTV algorithm we can determine the position of the particles compared to the focal planes of the cameras and determine their sizes more accurately using this information. 

To perform PTV, the most common setups include illumination by laser, but in this case the most appropriate illumination technique is back lighting with LED lamps. By using back lighting we can obtain defined contours of the observed particles that can be translated in to real size \cite{raffel1998particle,hagsater2008investigations,broder2007planar}. There are a few disadvantages of LED-back-illumination. Nonetheless, with an appropriate optical setup, these effects can be minimized. For example, LED light might saturate the images easily and without light-camera synchronization, blurry images will be obtained for particles that travel fast compared to the exposure times, this can be avoided if a very quick shutter speed is used. The quality of the optical system can be determined by a series of equations \cite[Chapter~2]{raffel1998particle}, where we can establish the minimum measurable particle size. 

From the images, the pixel lengths  for each particle in the vertical and horizontal axis of the image can be retrieved. From this information we can calculate the diameter or major and minor axis lengths on $x$ and $y$. Then, the pixel-to-world transformation --obtained in the calibration-- will be applied to retrieve a length estimation along $x$ and $y$. The accuracy of the length estimation will depend on the image resoultion and quality, the accuracy of the particle tracking and how well we can defined the optical properties of the system, which depend mainly in the camera and lens characteristics and the illumination. In the following, a description of the proposed methodology will be done as a proof of concept for the development of such measurements. 
 
\begin{figure}
	\includegraphics[width=0.60\textwidth]{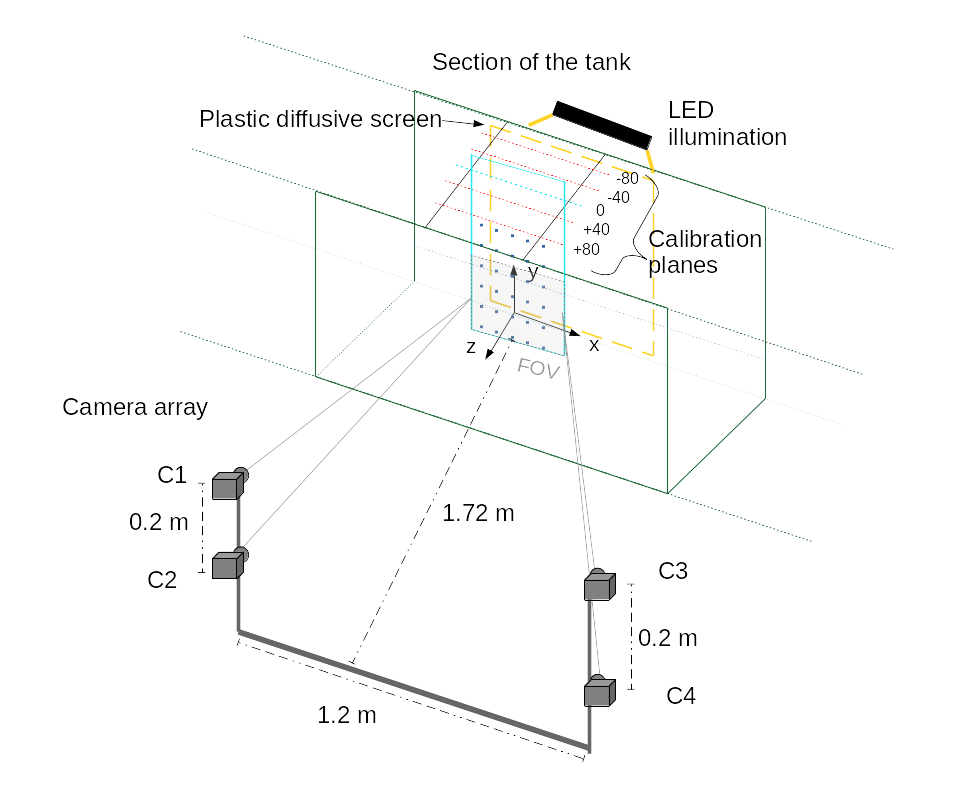}
	\includegraphics[width=0.40\textwidth]{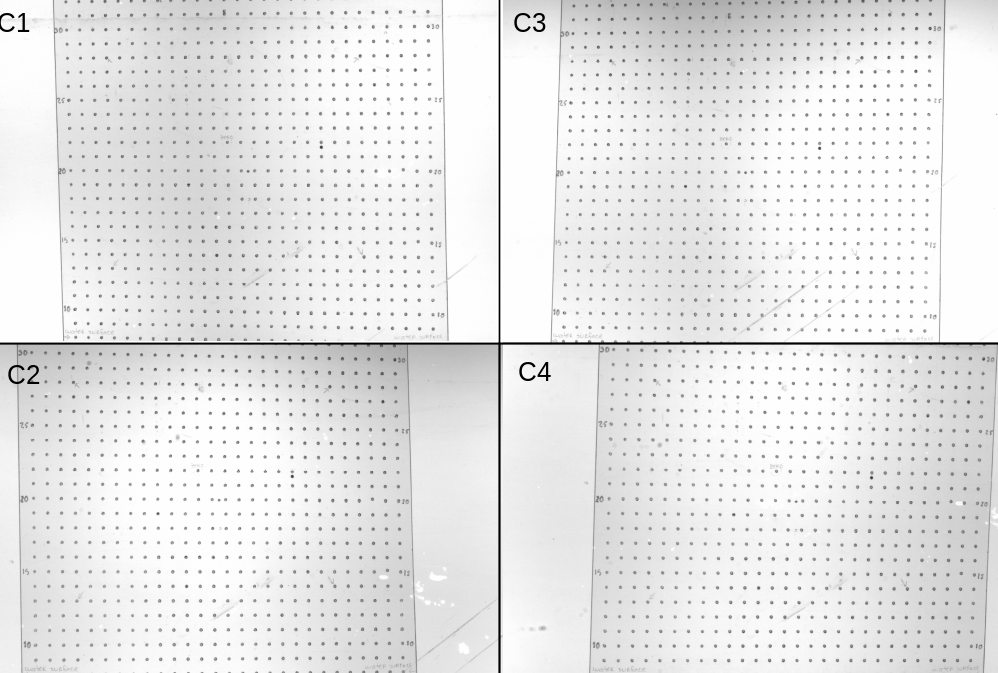}
	\caption{Experimental setup where the PTV measurements are performed, the image shows the array of cameras positioned to visualize approximately the same field of view. The multiple planes for the calibration with a flat target are also depicted, together with the coordinate system orientation, with $z=0$ in the center of the tank, $x$ parallel  to the tank length and $y$ in the vertical. To the right, and example of the obtained images of the flat target with the camera array.}
	\label{fig:setup}       
\end{figure}

\subsection{Experimental setup}
Figure \ref{fig:setup} shows a schematic description of the setup. This study has been conducted in the Hydrodynamics Laboratory of the University of Oslo. The setup is situated in a wave tank with dimensions $25\times0.52\times$ \m{1}. For the most part of this development, there has been no water in the tank, because the aim is to visualize water droplets traveling through air. To retrieve the images, four AOS Promon U1000 cameras were positioned in one side of the wave tank. The cameras had a resolution of $1920 \times 1080$ pixels and produced 8-bit gray scale images with a frame rate of 167 FPS. The positions of the cameras allowed them to capture a similar FOV. The illumination was provided by LED lamps in the opposite side of the tank and a white diffusive plastic sheet was used to distribute the light. According to equations ...  this optical system is reliable to measure particles with diameters $D_r > \mm{0.01}$ in a volume of approximately \mm{294}.   

To calibrate the PTV system, the method known as multiple plane calibration was used \cite{multiplane1,multiplane2}. This method consisted on taking images of a target in different planes of depth, i.e. moving the target in the $z$ coordinate. The target consisted on a flat acrylic plate with an array of visible holes, evenly spaced every \mm{10}. The positions of the holes in the target in reference to coordinate system $(x,y)$ are known, therefore by moving the target to different $z$, a swept or scan through the volume was made and the correlation between pixel location in each image with real space location was made by defining the epipolar geometry of the space \cite{maas1996contributions}.  

In the PTV analysis, it is common to use laser illumination and therefore images with dark background and light particles are produced. In contrast, images with light background and dark particles were created with this method, which allowed to obtain defined contours of the particles. To analize the images in the OpenPTV software, there are two possible paths. One is to modify the code to find light particles in dark background, which is possible with open source software, and the second option is to invert the gray scale of the images as to obtain light particles in dark background, then use this images directly in the code. While the first option could be more efficient, it also takes more time to adapt accurately to the code, and its implementation will require a careful development. Therefore, for this work we use the direct inversion of the images as to fit the requirements of the original code. So, the images that are analyzed by the PTV code appear to have light particles in a dark background.

\subsection{Determination of system parameters to estimate sizes in the analysis volume}
From the PTV algorithm, $n_{x}$ and $n_{y}$, the axis lengths in pixels of the different particles in $x$ and $y$, for each camera in the system, $C_i , i \in [1,2,3,4]$, along with the particles' positions can be obtained. With the proposed methodology we attempt to find a transformation:
\begin{equation}
T(n_{x},n_{y}; (x,y,z), \{C_{i}\}) = D,
\end{equation}
where $\{C_{i}\}$ are the parameters determined by the optical system, in particular each camera in the setup $C_i$, $(x,y,z)$ is the position of the particle and D is the diameter of the particle in the expected units (millimeter, meter or other). If the shape of the particles is assumed to be elliptical, we use the unique length measurement to identify particles sizes denominated as equivalent diameter, defined as $D_{e} = \sqrt{l_{x}l_{y}}$, where $l_{x}$ and $l_{y}$ are the axis lengths of the particle in $x$ and $y$ approximated as an ellipse in each image. Then, an average of $D_{e}$ is obtained for all the images. This averaged $D_e$ will be interpreted as the size of the particle. 

\begin{figure}
\includegraphics[width=0.50\textwidth]{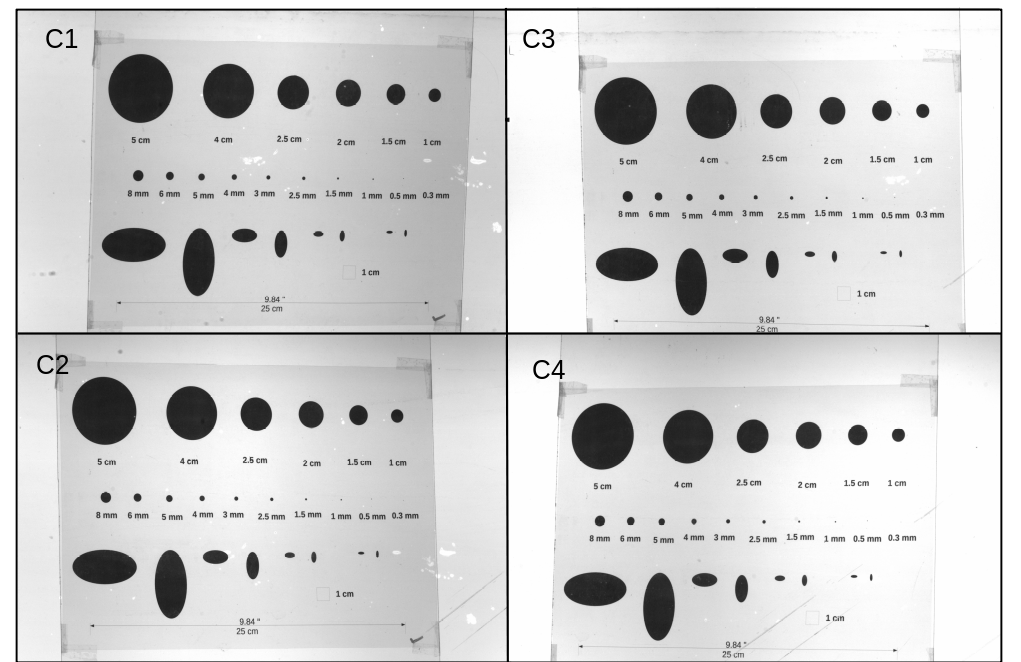}
\includegraphics[width=0.50\textwidth]{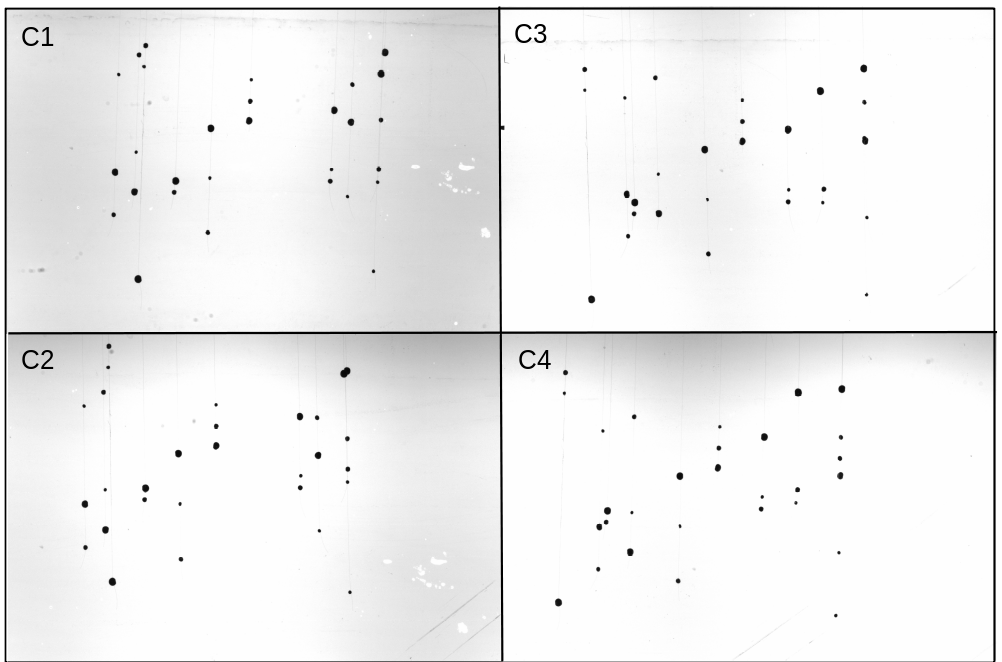}
\caption{Example of images taken with the camera array; to the right the flat target with circles and ellipses of different sizes is shown; to the left, the 3D particle array is shown.}
\label{fig:targets} 
\end{figure}

To obtain such transformation $T(n_{x},n_{y}: (x,y,z),\{C_{i}\})$,  the parameters for the cameras $\{C_{i}\}$ must be found by considering the effects of the optical system. Once the transformation  is defined, the estimation of the particle sizes can be made and compared to the real values, in such way the accuracy of the method can be estimated. A validation test was done to estimate the necessary parameters for the sizing methodology and understand the limits of such technique. The test consisted on using a flat target with circles and ellipses of different diameters from 0.3 mm to 50 mm (Fig. \ref{fig:targets}), the target was positioned in different depths, or $z$ values, and OpenPTV was used to detect the positions. Then an estimation of their sizes was made with the proposed procedure and compared to the known sizes. Finally the methodology is tested against a two different 3-dimensional arrays of particles with different sizes distributed over the volume of observation. In the following, the procedure to find some optical parameters will be described and the comparison of the results to the 3D particles will be shown. 

\subsection{Sources of error in the validation test}
\begin{figure}[h]
	\includegraphics[width=0.50\textwidth]{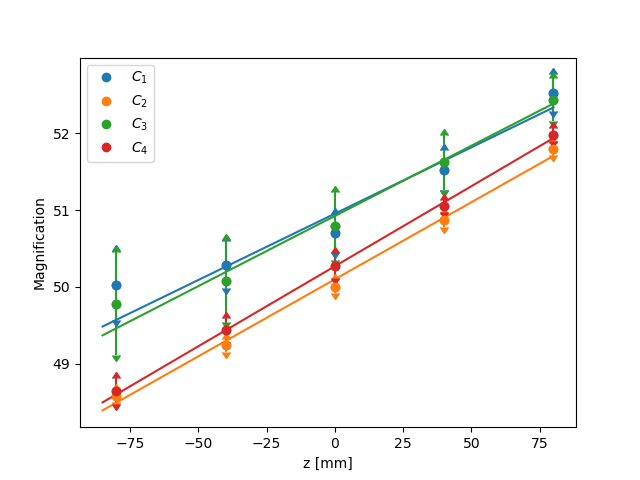}
	\includegraphics[width=0.50\textwidth]{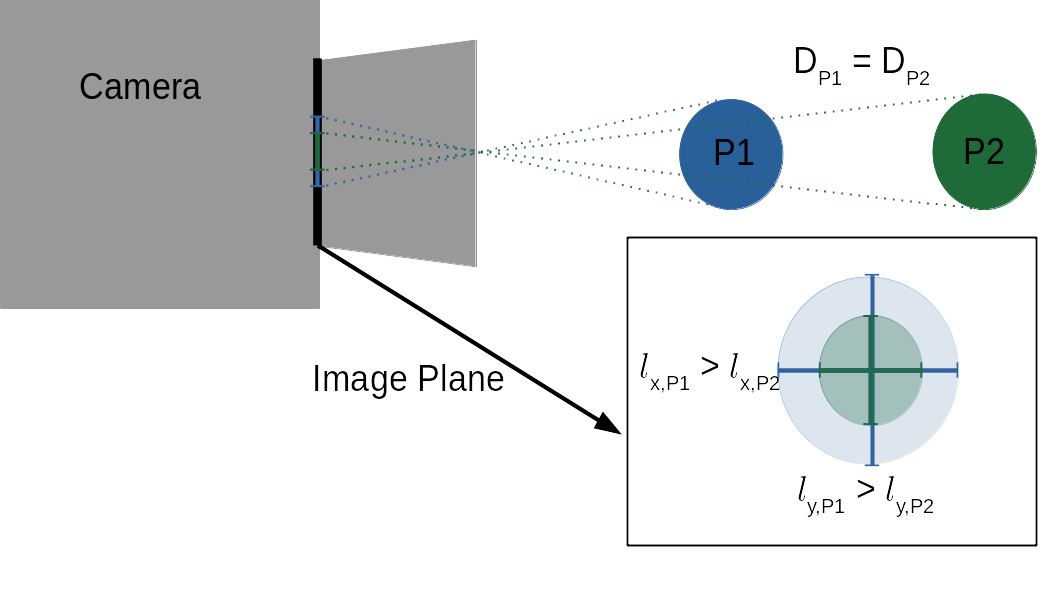}
	\caption{Left: $S_1$ described in eq. \ref{eq:s1}, the relation between magnification and depth $z$, calculated for each camera $C_i$, $i \in [1,2,3,4]$ in the array. The graph shows the magnification change with the depth (or z coordinate) of the calibration target. Right: a schematic description of the magnification in the image for two particles of the same diameter but with different $z$ coordinate.}
	\label{fig:pols-scale}
\end{figure}
The pixel-to-world transformation in the 3-dimensional space can be affected directly by the focal depth of the lenses, i.e. the magnification of the image depending on the distant to the lens. By using the calibration flat target with equidistant holes, the effects of the depth on the scale of the image can be estimated. The relation pixel size to real size depends on the position of the holes compared to the cameras, in particular the depth position or $z$ coordinate. The depth $z$ is approximated by the PTV algorithm and with the calibration, the magnification for each camera $C_i$ can be estimated as a function of $z$:
\begin{equation}\label{eq:s1}
    S_{1}(z,C_i) = \zeta_{1,i} z + \zeta_{0,i}.
\end{equation}
Linear fits as equation \ref{eq:s1} were fitted for $C_i$ in the different depths. The results are plotted in figure \ref{fig:pols-scale}. Then, images of the test target, with circles and ellipses, were obtained to obtain the axis lengths for the different sizes. 
Without loss of generality, if we choose a particluar camera $C_i$, the particle axis lengths in real units, $l_{x}$ and $l_{y}$, can be found as:
\begin{equation}
    l_{x}= \frac{n_{x}}{S_{1}(z,C_i)}, \quad l_{y}= \frac{n_{y}}{S_{1}(z,C_i)},
\end{equation}
where   $n_{x}, n_{y}$ are the number of pixels in the axis, $x$ and $y$ respectively, of the particle. Therefore, using the definition of $D_e$, the equivalent diameter of some particle $D_{p}$ can be described as:
\begin{equation}
    D_{p}= \sqrt{\frac{n_{x}}{S_{1}(z)} \frac{n_{y}}{S_{1}(z,C_i)}}=\frac{\sqrt{n_{x}n_{y}}}{S_{1}(z,C_i)} .
\end{equation}

\begin{figure}
	\includegraphics[width=0.60\textwidth]{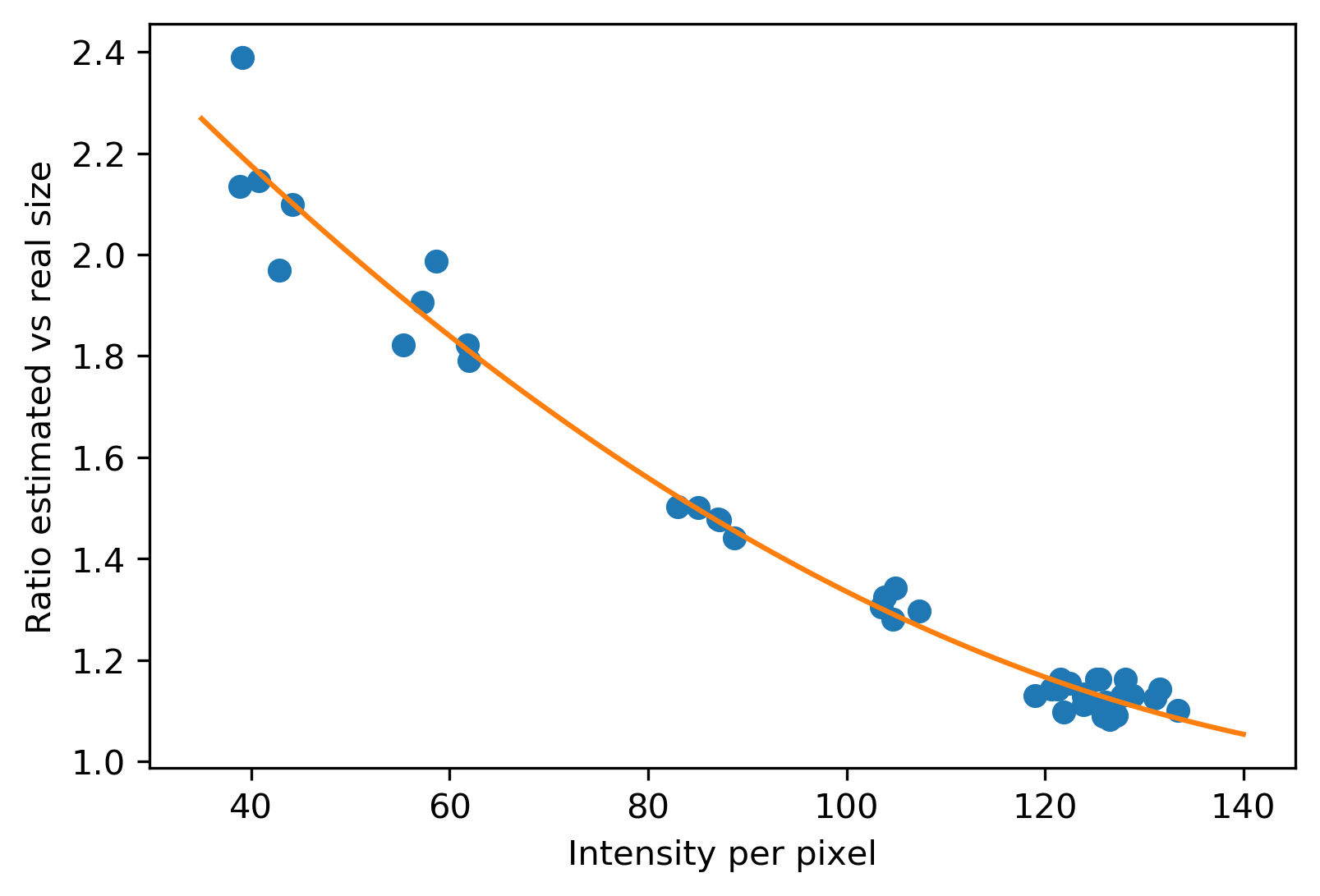}
	\caption{Second order polynomial for the gray value per pixel for particles with diameters below 22 pixels ($\approx 4.5 $mm). The graph shows the ratio between the estimated size and the real size, there is an over estimation of the size for particles with less gray value per pixel. The use of this polynomial can helps correct the estimation for particles with low contrast .}
	\label{fig:pol-gray} 
\end{figure}
\begin{figure*}
	\includegraphics[width=0.50\textwidth]{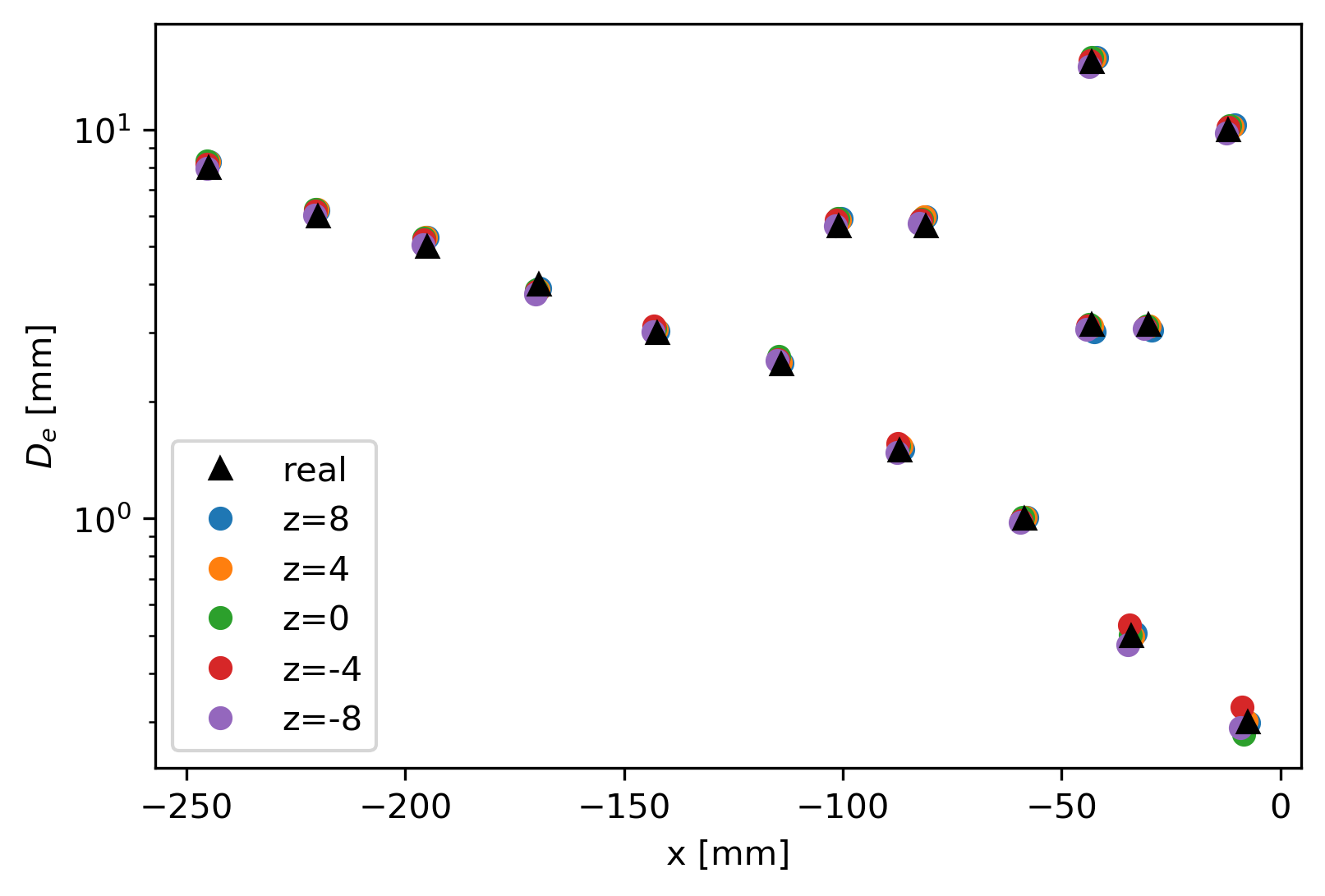}
	\includegraphics[width=0.50\textwidth]{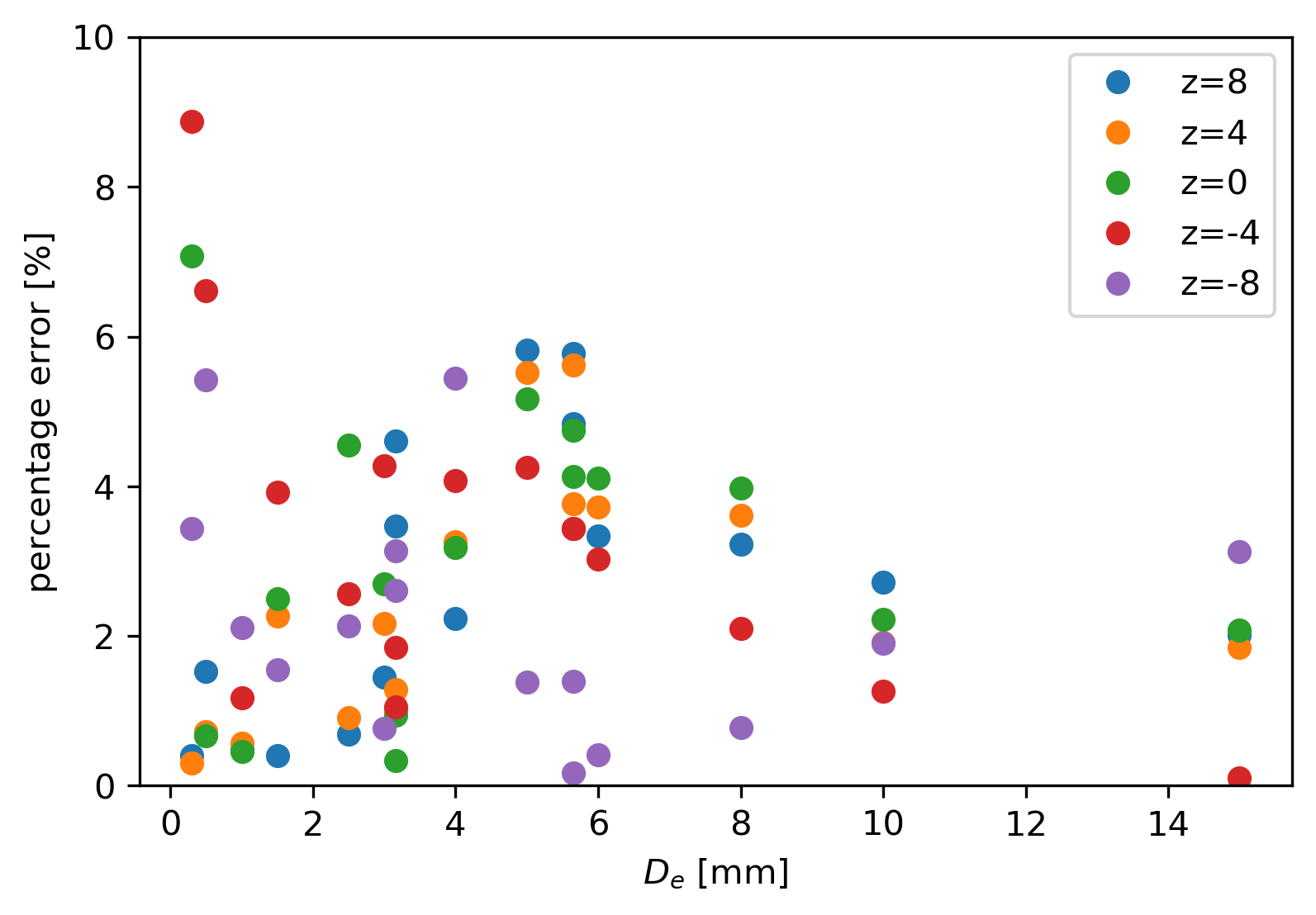}
	\caption{Results of the proposed size estimation applying the transformation $T(n_{x},n_{y}:z,g) $. To the  left, the estimated size compared to the real size of the different circles and ellipses in the target at different depths (z coordinate); to the right, the percentile error of this estimation compared to the real size.}
	\label{fig:circles}
\end{figure*}

In image processing, the particles are most commonly defined by the detection of blobs or edges in the image together with a threshold in the gray level or intensity. The particles can also be discriminated by pixel area and pixel length in the vertical and horizontal direction. These lengths and area are defined by the intensity or gray scale threshold, a value between 0 and 255 that has been fixed by the user during the image processing, which therefore has a direct impact in the approximated size of the particle. In addition, the further the particles are from the focal plane, the more blurry they get, this means that the detected diameter exceeds the expected value for a sharp-looking particle and a decrease of the mean intensity per pixel $I$ will also occur. A particle with diameter $D_{r}$ at a distance $z$ of the focal plane will have a detected diameter $D_p$ and mean intensity per pixel $I_p$, the value $I_p$ will be smaller than the value of $I$ for a sharp particle close to the focal plane, and the value $D_p$ will be larger than the value obtained for a sharp-looking particle of the same size. The ratio of the length estimation between a sharp-looking particle and a blurred particle can be expressed as $S_{2} = \frac{D_p}{D_{r}}$ and when plotting this ratio against $I$ of the detected particles, as shown in figure \ref{fig:pol-gray}, the correlation can be approximated by a second order polynomial:  
\begin{equation}
    S_{2}(I) = \alpha_{2} I^{2} + \alpha_{1} I + \alpha_{0},
\end{equation}
Then, the equivalent diameter of each particle can be estimated as:
\begin{equation}
    D_{e}=\frac{\sqrt{n_{x}n_{y}}}{S_{1}(z)S_{2}(I)}=T(n_{x},n_{y}:z,I),
\end{equation}
the transformation $T$ has been approximated, where the most relevant parameters have been the depth of the particle and its mean intensity per pixel. 
In figure \ref{fig:circles} we show the result of using the proposed transformation which gives percentage errors: $\frac{|D_{r}-D_{e}|}{D_{r}} \times 100 \leq 10\%$.  

\section{Validation on 3 dimensional particles}
\begin{figure}
  \includegraphics[width=0.5\textwidth]{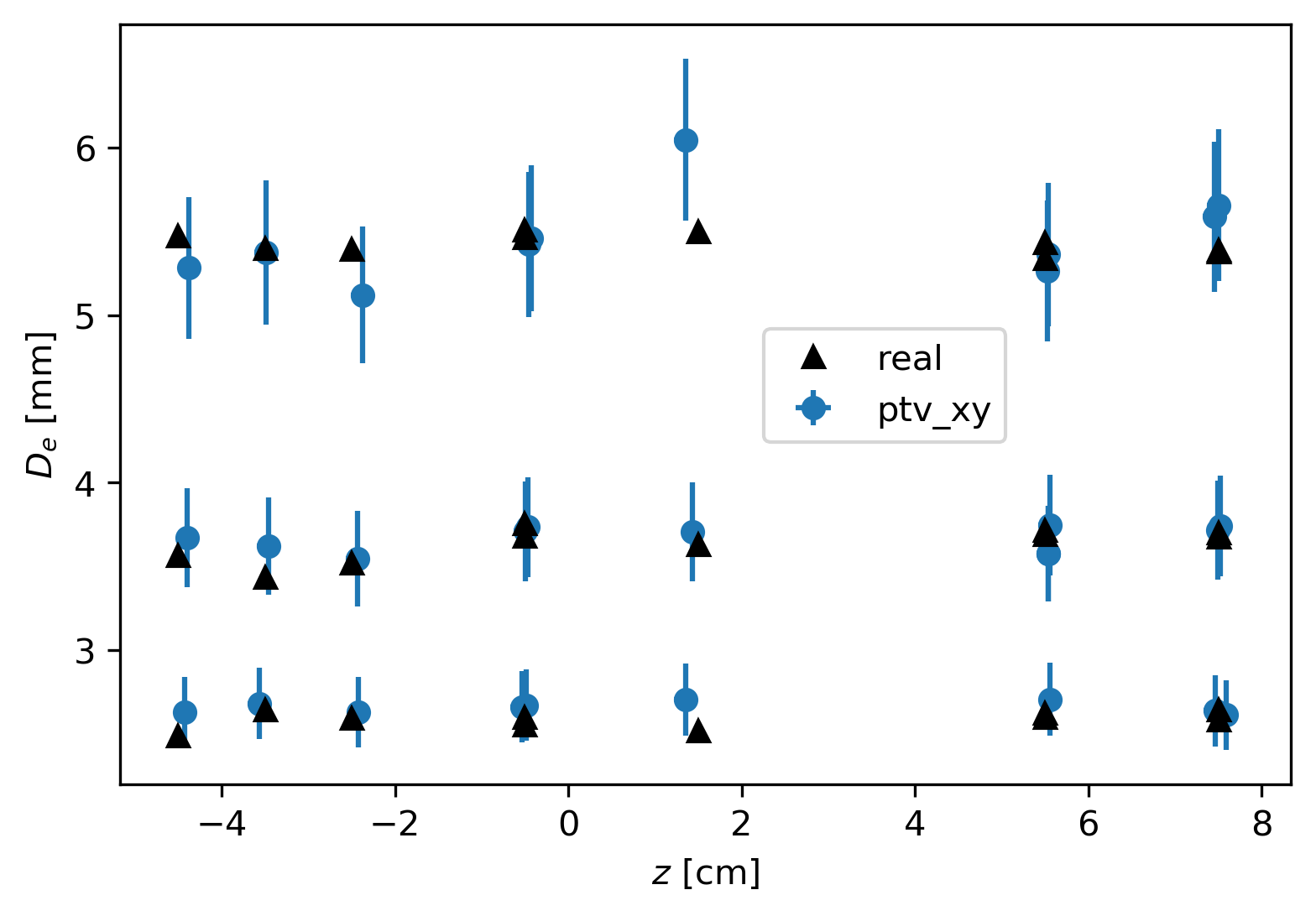}
  \includegraphics[width=0.5\textwidth]{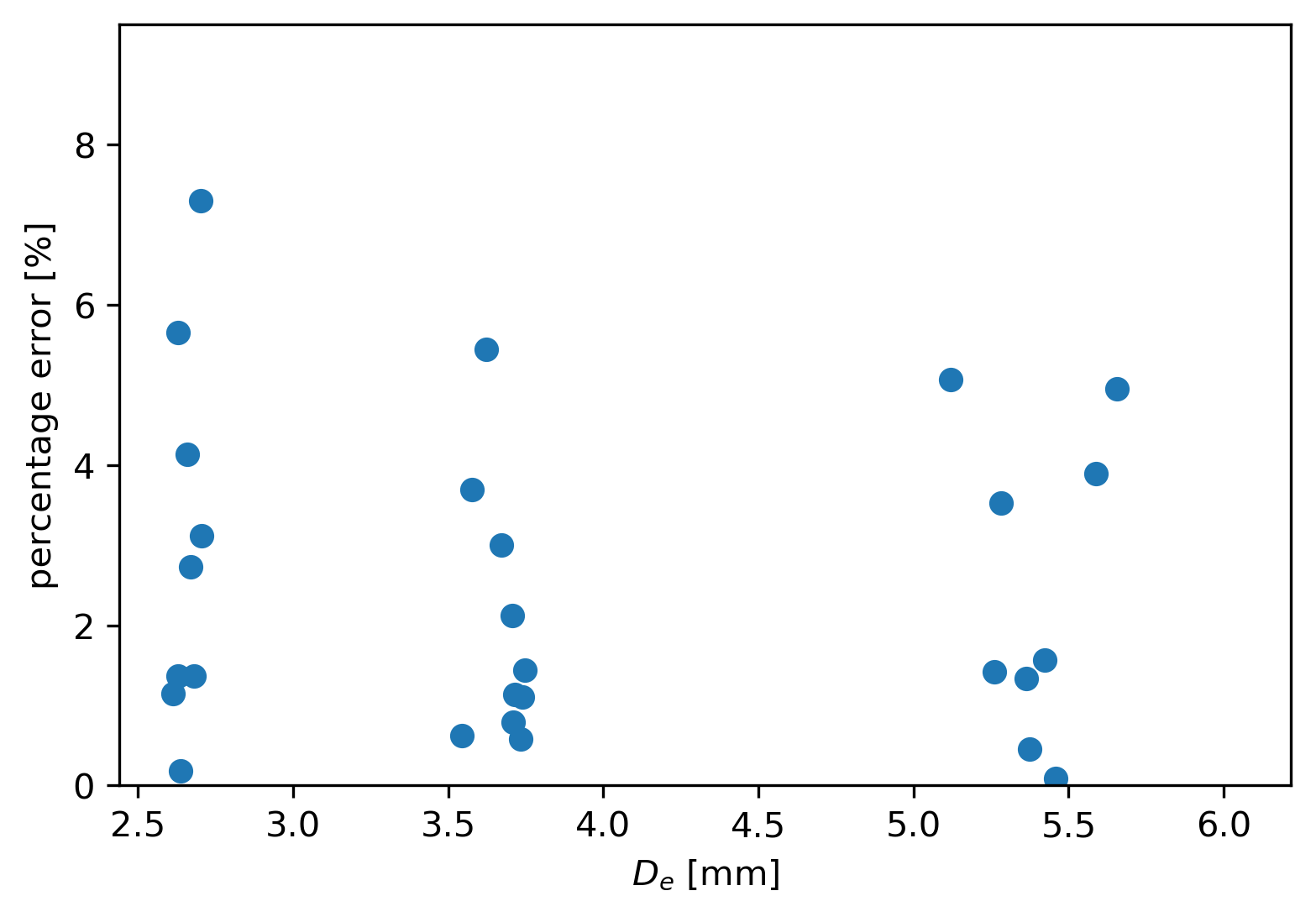}
\caption{Comparison between real sizes and sizes estimated by PTV for the array with 30 particles, with 3 sets of 10 particles with different sizes ($\approx 2.5, \quad \approx 3.5, \quad \approx 5.5$ mm for $D_e$). To the left, the size of the particle compared to the depth and to the right the percentage error for the different sizes. The error is no larger then 10\% in all cases. }
\label{fig:30parts}
\end{figure}
After obtaining the transformation using the flat target with circles and ellipses, we can test the new procedure on 3D particles. In this case, an array of solid particles was created with fishing lead weights of three different sizes. The dimensions of the particles were manually measured to make an equivalent diameter value for each of the particles. The three particle sizes are approximately  2.5,  3.5 and \mm{5.5}. Images obtained by the camera array for this particle array can be seen in figure \ref{fig:targets}. The comparison between the real sizes and the estimated sizes is presented in figure \ref{fig:30parts}, in the first plot we see the position of the particles in $z$ compared to the size. And in the second graph the percentage error of the size estimation is presented. The maximum error value is again 10\%, which shows that this first approximation to the sizing problem is reliable within certain limitations.  
\subsection{Other sources of error}
\begin{figure}
  \includegraphics[width=0.5\textwidth]{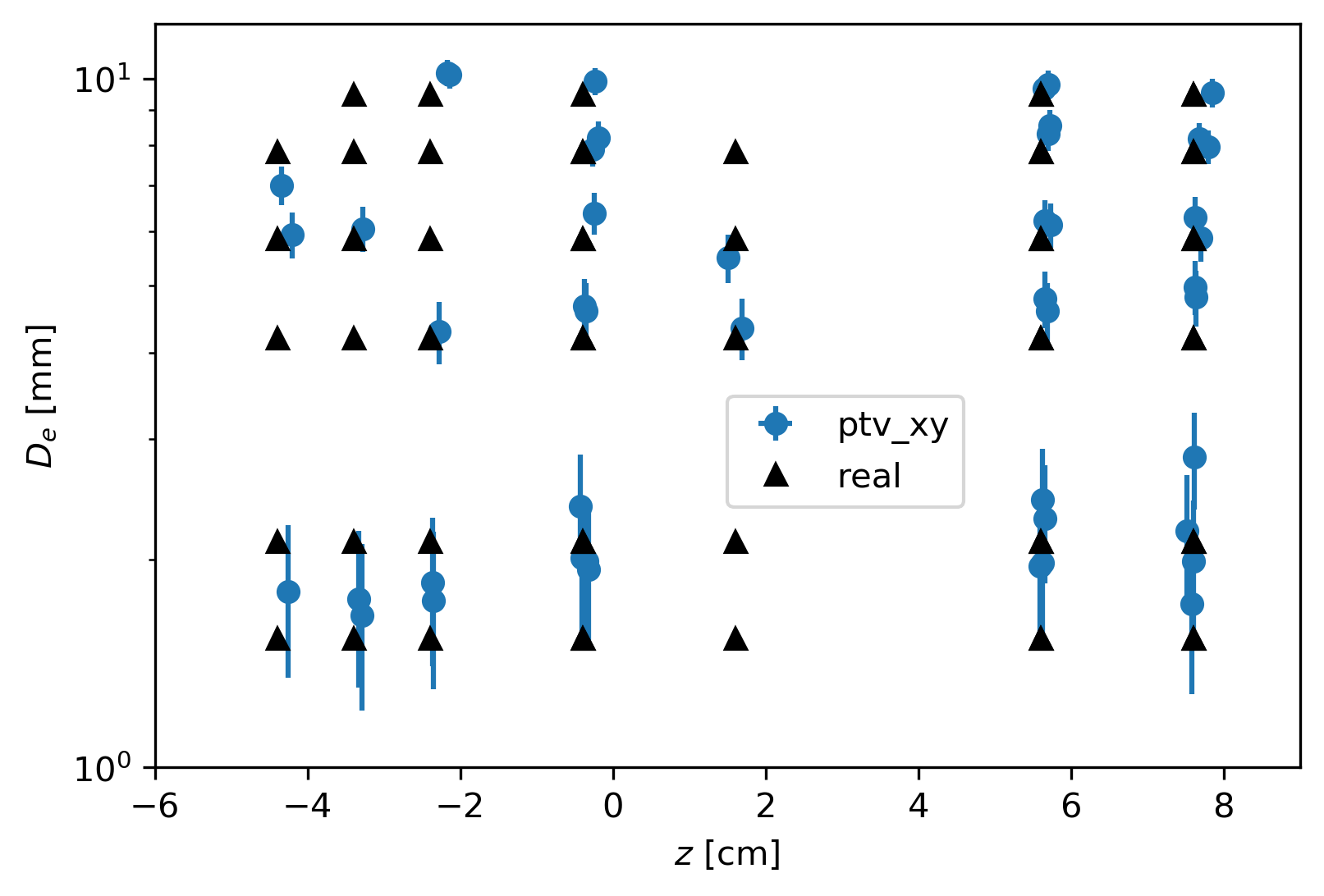}
  \includegraphics[width=0.5\textwidth]{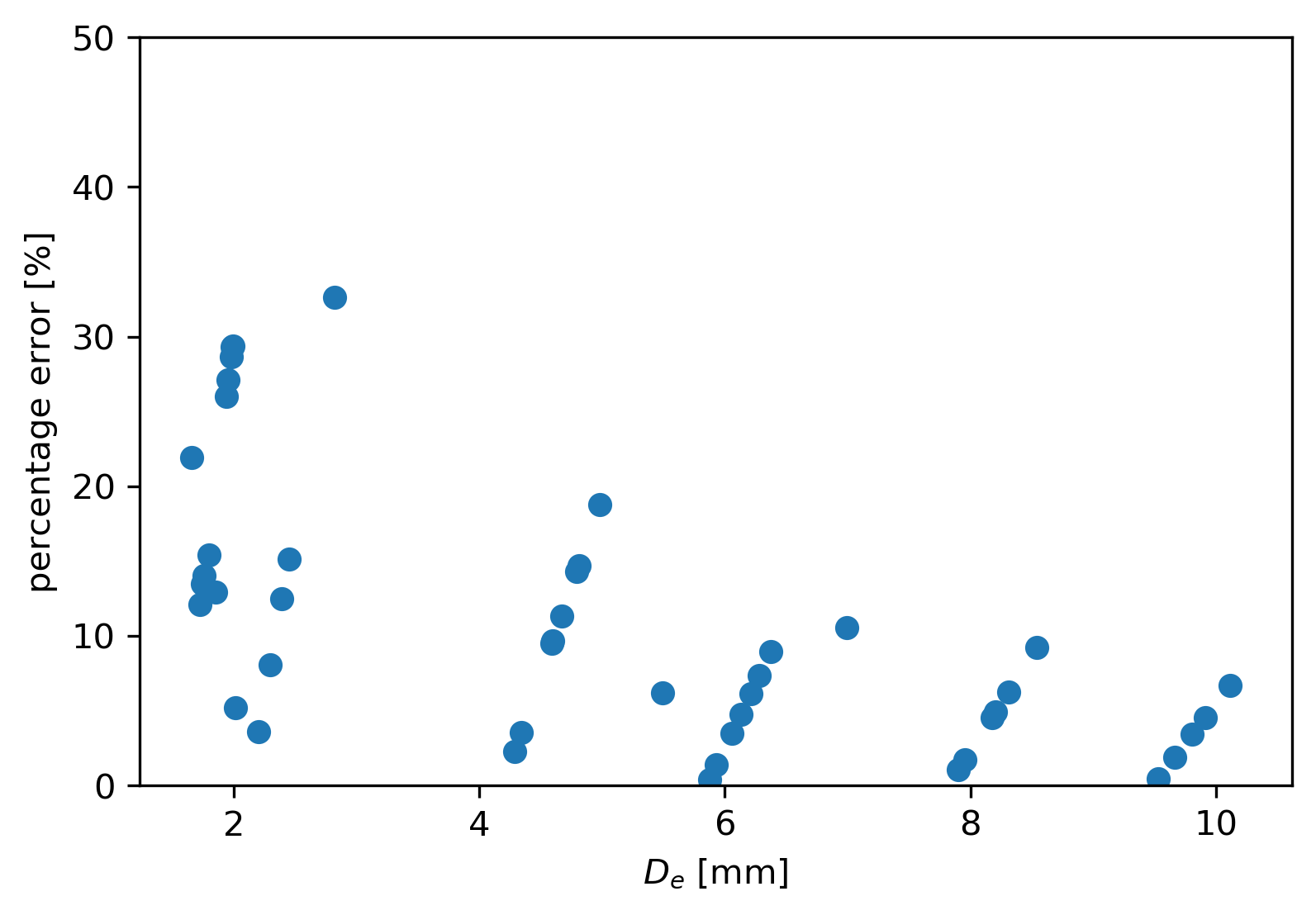}
\caption{Comparison between real sizes and sizes estimated by PTV for the array with 57 particles, with 6 sets of particles with different sizes ($\approx 1.5, \quad \approx 2.1, \quad \approx 4.2 ,\quad \approx 5.9, \quad \approx 7.8,$ and $\approx 9.5$ mm for $D_e$). To the left, the size of the particle compared to the depth and to the right the percentage error for the different sizes. The maximum error is 33\%, but all cases where the error is larger than 20\% correspond to small translucent particles.}
\label{fig:60parts}
\end{figure}
\begin{figure}
  \includegraphics[width=0.3\textwidth]{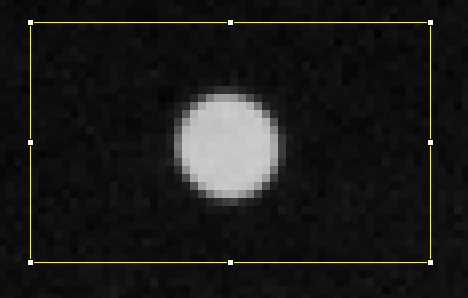}
  \includegraphics[width=0.31\textwidth]{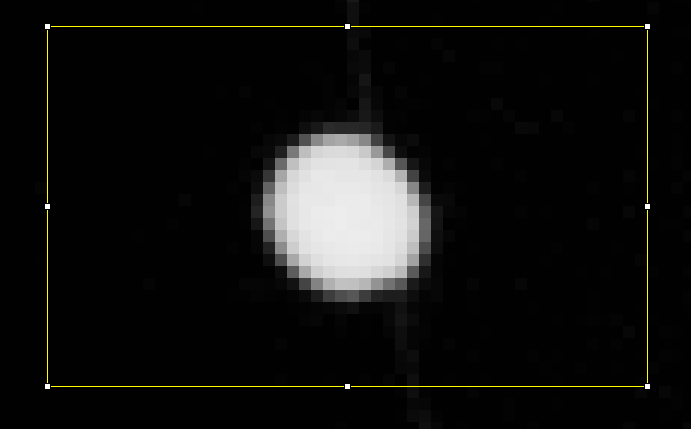}
  \includegraphics[width=0.3\textwidth]{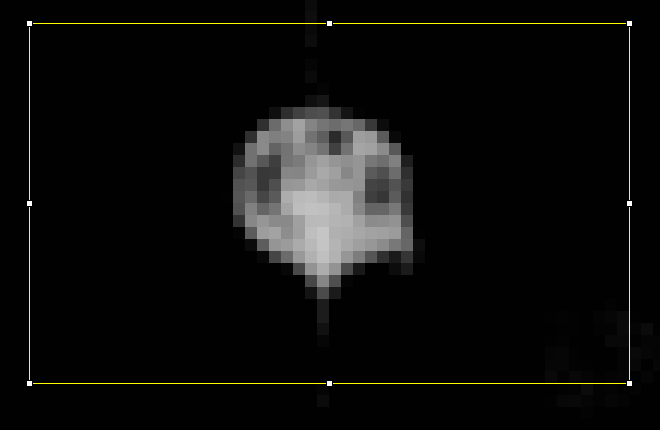}\\
  \includegraphics[width=0.3\textwidth]{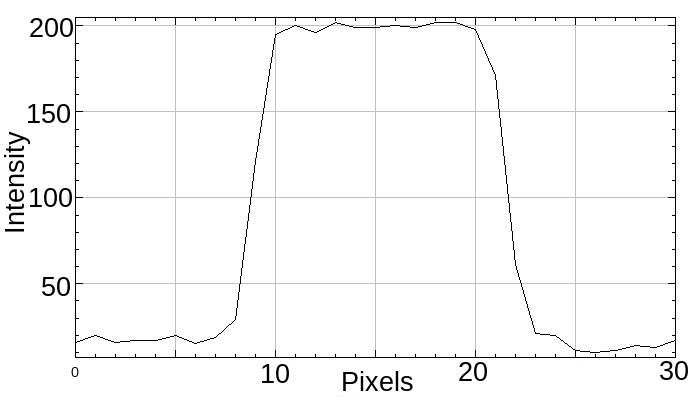}
  \includegraphics[width=0.3\textwidth]{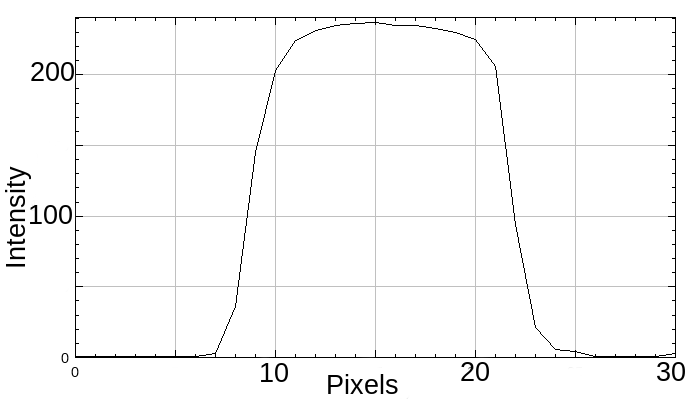}
  \includegraphics[width=0.3\textwidth]{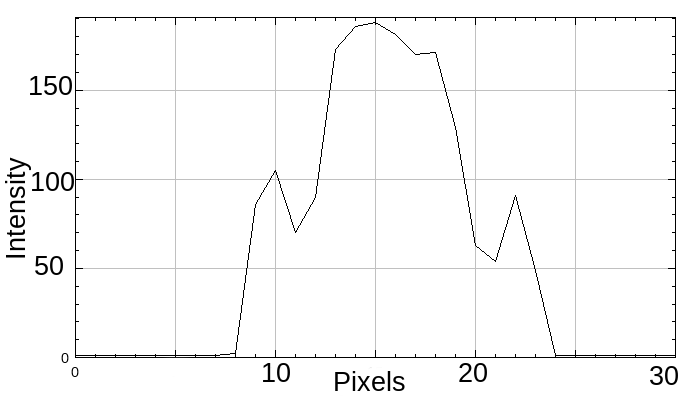}
\caption{Comparison between the two 3-D arrays (center and right) and the circle array (left) for a particle with diameter close to \mm{2.5}. The yellow line delimits an area of 30 by 50 pixels. The intensity plots corresponds to each case above and they show the gray value per pixel in the center line of the particle, which is equivalent to the pixel length of the particle in the horizontal direction.}
\label{fig:grays}
\end{figure}

Another 3-dimensional array of particles was made with plastic beads, some of them transparent and some others opaque. In total, 57 particles of 6 different sizes were distributed in the volume. The same procedure was applied to this array, and the results are presented in figure \ref{fig:60parts}. The graph shows that 85\% of the particles were tracked accurately by PTV and the maximum error in the size estimation is 35 \%.  The increase in the error is mainly on the particles with smallest sizes, which are the particles that are transparent. This shows that when the particles are not completely opaque to the light, the correction with the gray level will affect the estimation. Figure \ref{fig:grays} shows a comparison between the two 3-D arrays and the circle array for a particle (or circle) with diameter close to \mm{2.5}. In all cases, the particle has a similar length in pixels, but in the second 3-d array, where the particle is translucent, the intensity is lower, and therefore the gray value per pixel $g$ is lower. For a low $g$ the transformation $T$ assumes a larger correction shown in figure \ref{fig:pol-gray}, which in this case will lead to an underestimation of the size.

When obtaining the transformation, the assumption that all the visible particles will be opaque  to the light (as the circles) was implicit. In this case, introducing transparent or translucent particles will affect the accuracy of the method. Nonetheless, in the same way that opaque circles were used to obtain the transformation, an array of translucent circles or other features could be used while creating the transformation. Moreover, if the optical characteristics of the particles to analyze are known, the transformation and its correspondent array can be proposed in a way that it will account for these parameters. The main source of error will be the parameters that has not been predicted during the transformation's acquisition.  

\section{Concluding remarks}
\begin{figure}
  \includegraphics[width=\textwidth]{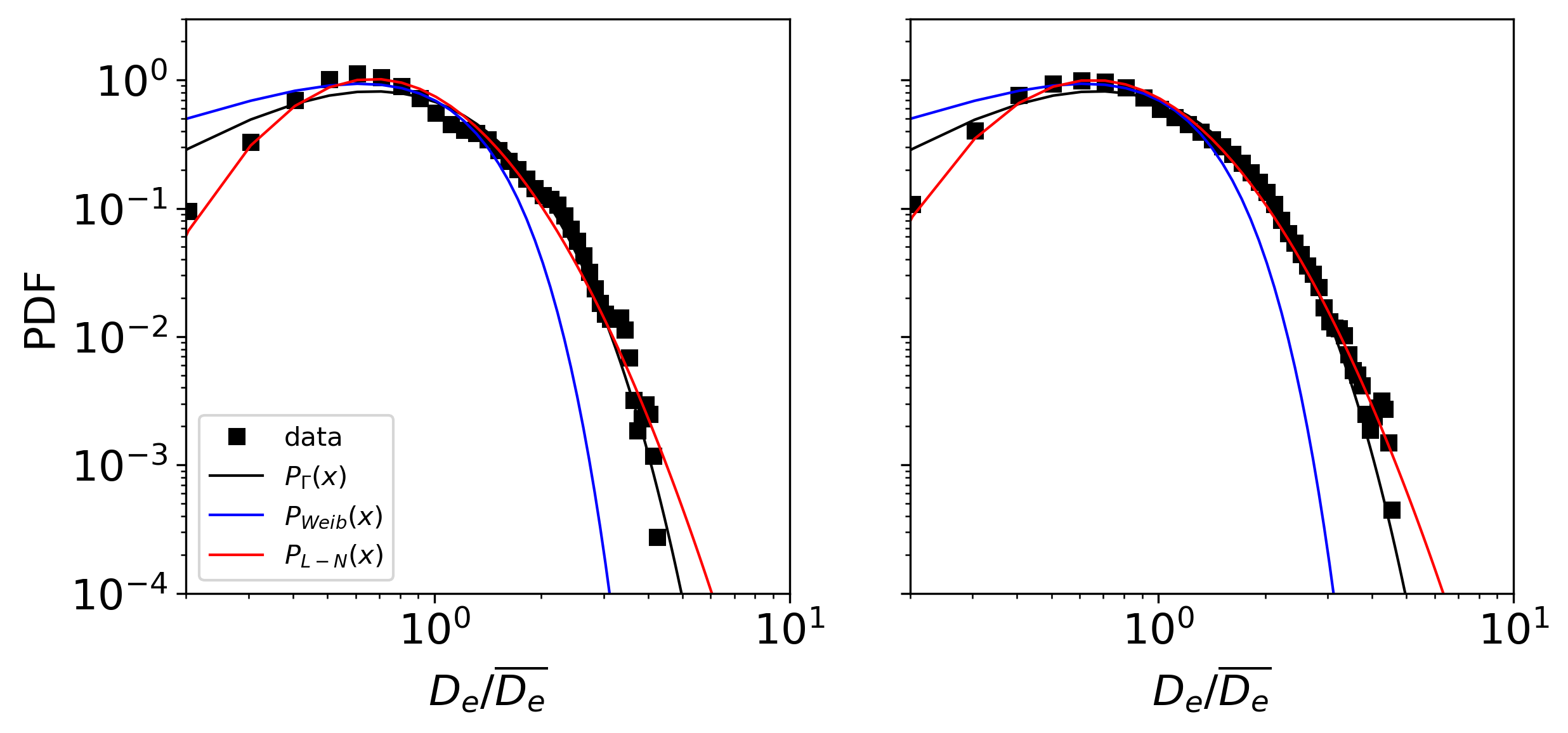}
\caption{Comparison of droplet size distribution obtained by the proposed method with other studies of droplet size distribution in similar cases. The square markers represent the experimental data and the solid lines show the distributions proposed by different studies: $P_{\Gamma}$ represents the droplet distribution proposed for a circular jet with wind forcing \cite{villermaux2004ligament}, $P_{Weib}$ represents the droplet distributions proposed for a droplet impact in a still pool \cite{roisman2006spray}, and $P_{L-N}$ represents the distribution proposed for droplet production after the impact of a monochromatic wave in a vertical wall \cite{watanabe2016size}. $D_e/\overline{D_e}$ is the equivalent diameter divided by the mean of the ensemble.}
\label{fig:waves}
\end{figure}

In this paper, we have presented a proof of concept to obtained particle size estimation assisted by a PTV algorithm that is open source. This has been a first approximation to the procedure which has proved to give large improvements by reducing the error of the estimation below 10\%. In the future, more sources of uncertainties can be identified and the estimation can be further improved. For example, the warping effects produced by the lens and the sensor can also be taken in to account to obtain a matrix of pixel-to-world scale for the whole field of view. With this methodology it is possible to perform meaningful analysis on 3D particle systems. This methodology has been applied in studies of droplet formation by breaking waves \cite{de2021experimental}. The results showed that the droplet size distribution estimated with this method was similar to those found by other studies of droplet size distribution as shown in figure \ref{fig:waves}. 

\bibliographystyle{spmpsci}      
\bibliography{bibliography.bib}   

\begin{thebibliography}{10}
\providecommand{\url}[1]{{#1}}
\providecommand{\urlprefix}{URL }
\expandafter\ifx\csname urlstyle\endcsname\relax
  \providecommand{\doi}[1]{DOI~\discretionary{}{}{}#1}\else
  \providecommand{\doi}{DOI~\discretionary{}{}{}\begingroup
  \urlstyle{rm}\Url}\fi

\bibitem{bachalo1984phase}
Bachalo, W., Houser, M.: Phase/doppler spray analyzer for simultaneous
  measurements of drop size and velocity distributions.
\newblock Optical Engineering \textbf{23}(5), 235583 (1984)

\bibitem{black1996laser}
Black, D.L., McQuay, M.Q., Bonin, M.P.: Laser-based techniques for
  particle-size measurement: a review of sizing methods and their industrial
  applications.
\newblock Progress in energy and combustion science \textbf{22}(3), 267--306
  (1996)

\bibitem{broder2007planar}
Br{\"o}der, D., Sommerfeld, M.: Planar shadow image velocimetry for the
  analysis of the hydrodynamics in bubbly flows.
\newblock Measurement Science and Technology \textbf{18}(8), 2513 (2007)

\bibitem{multiplane1}
del Castello, L.: Multi-plane calibration (2007).
\newblock
  \urlprefix\url{https://openptv-python.readthedocs.io/en/latest/add_doc.html}

\bibitem{openptv2014openptv}
Consortium, O., et~al.: Openptv (2014)

\bibitem{damaschke2002optical}
Damaschke, N., Nobach, H., Tropea, C.: Optical limits of particle concentration
  for multi-dimensional particle sizing techniques in fluid mechanics.
\newblock Experiments in fluids \textbf{32}(2), 143--152 (2002)

\bibitem{dracos1996particle}
Dracos, T.: Particle tracking in three-dimensional space.
\newblock In: Three-dimensional velocity and vorticity measuring and image
  analysis techniques, pp. 209--227. Springer (1996)

\bibitem{multiplane2}
Goumnerov, H.: Openptv. installation and user manual. (2007).
\newblock
  \urlprefix\url{https://openptv-python.readthedocs.io/en/latest/add_doc.html}

\bibitem{hagsater2008investigations}
Hags{\"a}ter, S., Westergaard, C., Bruus, H., Kutter, J.: Investigations on led
  illumination for micro-piv including a novel front-lit configuration.
\newblock Experiments in fluids \textbf{44}(2), 211--219 (2008)

\bibitem{katz2010applications}
Katz, J., Sheng, J.: Applications of holography in fluid mechanics and particle
  dynamics.
\newblock Annual Review of Fluid Mechanics \textbf{42}, 531--555 (2010)

\bibitem{koothur2021tracking}
Koothur, V.: Tracking and sizing of particles in the mie scattering regime
  using a laser scanning technique  (2021)

\bibitem{kozul2019scanning}
Kozul, M., Koothur, V., Worth, N.A., Dawson, J.R.: A scanning particle tracking
  velocimetry technique for high-reynolds number turbulent flows.
\newblock Experiments in Fluids \textbf{60}(8), 1--14 (2019)

\bibitem{luthi2005lagrangian}
L{\"u}thi, B., Tsinober, A., Kinzelbach, W.: Lagrangian measurement of
  vorticity dynamics in turbulent flow.
\newblock Journal of Fluid mechanics \textbf{528}, 87--118 (2005)

\bibitem{maas1993particle}
Maas, H., Gruen, A., Papantoniou, D.: Particle tracking velocimetry in
  three-dimensional flows.
\newblock Experiments in fluids \textbf{15}(2), 133--146 (1993)

\bibitem{maas1996contributions}
Maas, H.G.: Contributions of digital photogrammetry to 3-d ptv.
\newblock In: Three-dimensional velocity and vorticity measuring and image
  analysis techniques, pp. 191--207. Springer (1996)

\bibitem{malik1993particle}
Malik, N., Dracos, T., Papantoniou, D.: Particle tracking velocimetry in
  three-dimensional flows.
\newblock Experiments in fluids \textbf{15}(4), 279--294 (1993)

\bibitem{raffel1998particle}
Raffel, M., Willert, C.E., Kompenhans, J., et~al.: Particle image velocimetry:
  a practical guide, vol.~2.
\newblock Springer (1998)

\bibitem{roisman2006spray}
Roisman, I.V., Horvat, K., Tropea, C.: Spray impact: rim transverse instability
  initiating fingering and splash, and description of a secondary spray.
\newblock Physics of Fluids \textbf{18}(10), 102104 (2006)

\bibitem{de2021experimental}
de~la Torre, R., Vollestad, P., Jensen, A.: Experimental investigation of
  droplet generation by post-breaking plunger waves.
\newblock arXiv preprint arXiv:2112.01396  (2021)

\bibitem{tropea2011optical}
Tropea, C.: Optical particle characterization in flows.
\newblock Annual Review of Fluid Mechanics \textbf{43}, 399--426 (2011)

\bibitem{villermaux2004ligament}
Villermaux, E., Marmottant, P., Duplat, J.: Ligament-mediated spray formation.
\newblock Physical review letters \textbf{92}(7), 074501 (2004)

\bibitem{watanabe2016size}
Watanabe, Y., Ingram, D.: Size distributions of sprays produced by violent wave
  impacts on vertical sea walls.
\newblock Proceedings of the Royal Society A: Mathematical, Physical and
  Engineering Sciences \textbf{472}(2194), 20160423 (2016)

\end{thebibliography}

\end{document}